\documentclass[aps,twocolumn,showpacs,prl,10pt]{revtex4-1}

\usepackage{amsmath}
\usepackage{amssymb}
\usepackage{amsthm}
\usepackage{graphicx}
\usepackage{epsfig}

\newcommand{\sech}{{\rm sech\,}}
\newcommand{\R}{\mathbb{R}}

\begin{document}

\title{Nonlinear supratransmission in multicomponent systems}
\author{P. Anghel-Vasilescu, J. Dorignac, F. Geniet, J. Leon, 
M. Taki}\altaffiliation{Permanent address: Laboratoire de Physique des Lasers,
Atomes et Mol\'ecules, CNRS-INP-UMR8523,Universit\'e des Sciences et
Technologies de Lille, 59655 Villeneuve d'Ascq, (France)}
\affiliation{Laboratoire de Physique Th\'eorique et Astroparticules \\
CNRS-IN2P3-UMR5207, Universit\'e Montpellier 2, 34095 Montpellier (France)}

\begin{abstract}
A method is proposed to solve the challenging problem of determining the
supratransmission threshold (onset of instability of harmonic boundary driving
inside a band gap) in multicomponent nonintegrable nonlinear systems. It is
successfully applied to the degenerate three-wave resonant interaction in a
birefringent quadratic medium where the process generates spatial gap solitons.
No analytic expression is known for this model showing the
broad applicability of the method to nonlinear systems.
\end{abstract}

\pacs{O5.45.Yv, 42.65.Tg\hfill Phys. Rev. Lett. 105  (2010) 074101}

\maketitle

\paragraph*{Introduction.}

Nonlinear supratransmission (NST) in a medium possessing a natural forbidden
band gap is a process by which nonlinear structures, gap solitons, are
generated by an applied periodic boundary condition at a frequency in the band
gap. Discovered in the pendula chain (sine-Gordon model) \cite{nst-prl}, and
further studied for fully discrete chain in Refs. \cite{aubry} and \cite{daz},
it has been applied, among others, in Bragg media (coupled mode equations in
Kerr regime) \cite{nst-bragg} allowing to explain the experiments of Ref.
\cite{taverner}, and also to coupled-wave-guide arrays (nonlinear Schr\"odinger
model) \cite{ramaz1}\cite{jl}. Nonlinear supratransmission results from an
instability of the evanescent
wave profile created by the driving \cite{instab}\cite{susanto} that manifests
itself above a threshold amplitude. Today this threshold has been obtained in
single component systems by making use of the explicit solution of the model
equation and seeking its maximum allowed amplitude at the boundary. 

Predicting the threshold value is of fundamental importance for physical
applications such as soliton generation or conception of ultrasensitive
detectors. Indeed, on the one side NST is a very efficient means to generate gap
solitons: while an incident single pulse with carrier wave at forbidden
frequency would be mainly reflected, an incident continuous wave excitation
easily produces gap soliton as experimentally shown in Bragg media
\cite{taverner}.  On the other side such systems seeded by a CW excitation
slightly below the threshold will be extremely sensitive to any applied signal,
detected either through generation of gap solitons or by bistable behavior, see
e.g. \cite{ramaz2}. 

We address in this Letter the practical question of evaluating NST thresholds in
multicomponent systems (where the instability of either wave separately can
induce soliton formation in all channels), and moreover when the system has no
explicit solution allowing for threshold prediction. This is the case with
second harmonic generation in a birefringent medium with quadratic nonlinearity
\cite{buryak}. The model is a two-component system which does not possess
analytic expression of solitonlike solutions and which is a key model in order
to study their existence and stability; see, e.g., \cite{majid}. 

We shall develop a method based on an asymptotic solution  obtained by
asymptotic series expansion, which provides an accurate NST threshold
prediction. As NST requires driving in the forbidden band, the linear evanescent
wave is the natural keystone upon which to build the series. The method is
restricted  to neither the specific case of second harmonic generation nor to
the quadratic nature of the nonlinearity. Moreover, it can be applied to a wide
class of nonintegrable multicomponent nonlinear systems since it does not
require known analytical expressions for their solutions. The method thus
furnishes a practical tool highly interesting for further applications in any
multi-component coupled-wave system.

\paragraph*{Birefringent gap solitons.}
\begin{figure}[ht] 
\centerline{\epsfig{file=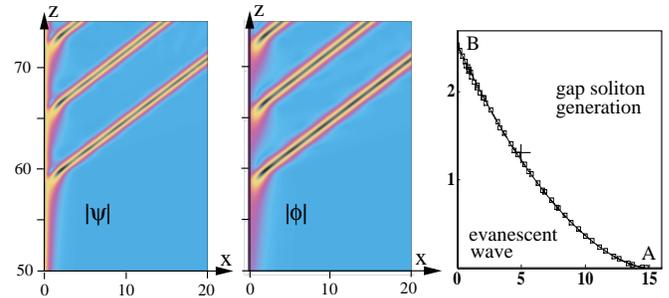,width=\linewidth}}
\caption{(color online) Left: intensity plots of a typical numerical simulation
of the D3W system (\ref{D3W}) with $\alpha=1.1$ subject to boundary conditions
(\ref{D3W-bound}) with $A=5$, $B=1.3$ [the cross on the $(A,B)$ plot]. The
maximum amplitudes of the emitted soliton are $|\psi|_m\sim 4.5$ and
$|\phi|_m\sim 5.5$. Right: threshold curve in the $(A,B)$ plane where dots
result from numerical simulations \cite{note-simuls} and the curve is given by
the solution of (\ref{BGS_NST_condition}).}\label{fig:1}
\end{figure}
Let us consider a birefringent medium in permanent regime, namely assuming
perfect
frequency matching. In that case,  degenerate spatial three-wave model
(D3W) reduces  to \cite{buryak}\cite{majid}
\begin{align}\label{D3W}
 & i\partial_z \psi+\frac{\alpha }{2}\partial_x^2 \psi-\delta
\psi+\phi^2=0,\nonumber\\
 & i\partial_z \phi+\partial_x^2 \phi+\psi\phi^*=0.
\end{align}
where $\phi(x,z)$ [respectively  $\psi(x,z) e^{i\delta z}$] is the scaled static
envelope of the signal wave with frequency $\omega$ and wave number $k$
[respectively second harmonic at frequency $\omega'=2\omega$ and wave number
$k'$] and $\delta$ is the missmatch wavenumber in the propagation direction $z$
defined by $k'=2k-\delta$. Last, $x$ is the transverse direction and
$\alpha=1+\delta/k'$ by definition. The system (\ref{D3W}) is subject to the
boundary condition
\begin{equation}\label{D3W-bound}       
 \psi(0,z)=Ae^{2iz},\quad \phi(0,z)=Be^{iz}
\end{equation}
on the strip $x>0$, $z\in[0,L]$, with vanishing conditions as $x\to\infty$  and
now with normalization $k'=2$ (that is $\alpha=1+\delta/2$) \cite{note-BGS}. 

For the sake of simplicity, we only consider here the boundary conditions
(\ref{D3W-bound}), although the wave number may in general be chosen different
from $1$. We see on Fig.\ref{fig:1} the generation, through the evanescent
coupling, of birefringent gap solitons (BGS) formation and propagation
above a threshold curve in the amplitude plane $(A,B)$. It is worth
pointing out that the present situation is fundamentally different from studies
of initial-value problems where field values are given at $z=0$, as,
e.g., in the theoretical prediction \cite{leo-assanto} and experimental
realization \cite{canva} for nondegenerate 3-wave interaction.

\paragraph*{Asymptotic series solution.}

Given the boundary values (\ref{D3W-bound}), we seek stationary solutions of
the form
\begin{equation}\label{psi-phi}
 \psi(x,z)=u(x)e^{2iz},\quad \phi(x,z)=\sqrt{\alpha}\, v(x)e^{iz},
\end{equation}
with real-valued functions $u$ and $v$ vanishing at infinity
(the factor $\sqrt{\alpha}$ has been included for convenience). The system
(\ref{D3W}), now with $\delta=2(\alpha-1)$,
provides for $u$ and $v$ the following parameter free equations
\begin{equation} \label{eq-uv}
\partial_x^2 u- 4 u + 2 v^2 = 0,  \quad
\partial_x^2 v - v + u v = 0,
\end{equation} 
Treating nonlinear terms as perturbative, we first solve the linearized equation
for $v$ and obtain $v=\beta e^{-x}$. Substituting this result in the equation
for $u$, we find that the $v^2$ term is resonant and generates a solution of the
form $u=(\mu + \beta^2x/2)e^{-2x}$ where $\mu$ and $\beta$ are two arbitrary
constants. This is the general solution of the quasilinear system $\partial_x^2
u- 4 u + 2 v^2 = 0$, $\partial_x^2 v=v$, that vanishes as $x\to\infty$.

The structure of (\ref{eq-uv}) now indicates that $u$ and
$v$ may be expressed as the following asymptotic series
\begin{align} \label{series-uv}
&u=\sum_{n=0}^{\infty} P_n(x)e^{-2(n+1)x},\quad v=\sum_{n=0}^{\infty}
Q_n(x)e^{-(2n+1)x},\nonumber\\
& Q_0(x)=\beta,\quad P_0(x)=\mu+\beta^2 x/2.
\end{align} 
By inspection, the polynomials $P_n(x)$ and $Q_n(x)$ are of degree $n$ (except
$P_0$ of degree 1) and obey a system of differential-recurrence equations
obtained by replacing \eqref{series-uv} in \eqref{eq-uv}. Their coefficients are
then recursively given in terms of the two independent parameters $\beta$ and
$\mu$. To ensure an accurate determination of the threshold curve, we have
evaluated the first 17 terms  of the series (\ref{series-uv}), which takes up to
a minute on a PC computer (with  MAPLE or MATHEMATICA). The first ones are e.g.
$Q_1=-3\beta^3/64-\mu\beta/8-\beta^3 x/16$ and 
$P_1=\beta ^2 (24 \mu +17 \beta^2)/576+x \beta^4/48$.
The advantage of this method is to be applicable to  any system driven in a
forbidden band, which is essential when no soliton solution is known. Notice
that, in the absence of resonant terms in the equations, the polynomials
involved in the asymptotic series would simply be constants as e.g. in the
Manakov system below. 

\paragraph*{NST threshold prediction.}

Once the series (\ref{series-uv}) has been determined up to a given truncation
order $N$, imposing the boundary conditions $u(0)=A$ and $v(0)=B$ leads to the
two driving amplitudes $A(\beta,\mu)$ and $B(\beta,\mu)$ explicitly expressed
in terms of $\beta$ and $\mu$. Assuming that the supratransmission threshold
curve is given by the maximum value of one of those, the other one being held
constant, we can use a Lagrange parameter $\lambda$ and write the extremum
condition as $\partial_{\mu}A-\lambda\partial_{\mu}B =0$ and
$\partial_{\beta}A-\lambda\partial_{\beta}B=0$. This finally leads to the
condition of vanishing Jacobian
 \begin{equation} \label{BGS_NST_condition}
J(\beta,\mu)=(\partial_{\beta}A) (\partial_{\mu}B)
-(\partial_{\mu}A) (\partial_{\beta}B)  = 0.
\end{equation} 
With a high enough truncation order ($N=16$ here), the threshold curve is best
obtained as the zero contour of the surface $J(\beta,\mu)$ plotted
parametrically as a function of the amplitudes $A(\beta,\mu)$ and
$B(\beta,\mu)$, as  presented in Fig.\ref{fig:1}.

The condition (\ref{BGS_NST_condition}) is symmetric with respect to  $A$ and
$B$, thus there is no need to specify which maximum amplitude is sought.
Moreover condition (\ref{BGS_NST_condition}) does not depend on the specific
choice of parameters, provided they are independent. For instance, using the 
new  parameters $\eta$ and $\sigma$ defined by 
\begin{equation}\label{new-par}
\beta=\exp(\eta),\quad \mu=(\sigma-\eta)\exp(2\eta),
\end{equation}
the NST threshold condition becomes $J(\sigma,\eta)=0$. The interesting point
here is that for fixed $\sigma$, the new parameter $\eta$ operates a shift of
the solutions (\ref{series-uv}) $u(x)\to u(x-\eta)$ and $v(x)\to v(x-\eta)$. 
Such a translation parameter always exists in systems that possess a translation
invariance on the whole $x$-axis. 

\paragraph*{Generalizations.}

Generalization of the procedure to $M$-component systems is straightforward. To
this end, let us denote the components by $\phi_m(x,z)=u_m(x)e^{i\nu_m
z}$ and their amplitudes by $u_m(0)=A_m(\{\eta_n\})$, where $\eta_n$ are the
$M$ parameters of the solution (e.g. in the asymptotic solution). 
Finding the NST threshold manifold amounts to setting to zero the determinant
of the Jacobi matrix of the amplitudes with respect to the parameters, that is
\begin{equation} \label{General_NST_condition}
{\rm det}\,[J_{mn}]  = 0,\quad J_{mn} = 
\frac{\partial A_m}{\partial \eta_n}.
\end{equation} 

\begin{figure}[ht] 
\centerline{\epsfig{file=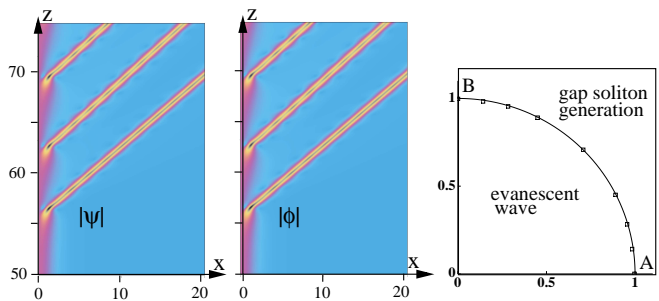,width=\linewidth}}
\caption{(color online) Left: typical numerical simulation of
(\ref{manakov-sys}) submitted to boundary conditions (\ref{bound-manakov}) with
$A=0.45$ and $B=0.9$. The maximum amplitudes of the emitted soliton are
$|\psi|_m\sim 1.2$ and $|\phi|_m\sim 2.3$. Right: threshold curve in the
$(A,B)$-plane where the dots result from numerical simulations
\cite{note-simuls} and the curve is the circle $A^2+B^2=1$.}\label{fig:2}
\end{figure}

To illustrate our result with another interesting
problem, we may consider a
multicomponent system having soliton solutions, the (integrable) Manakov system
\cite{manakov}, written here for spatial fields as
\begin{align}\label{manakov-sys}
 & i\partial_z \psi+\partial^2_x \psi+2(|\phi|^2+|\psi|^2)\psi=0,\nonumber\\
 & i\partial_z\phi+\partial^2_x \phi+2(|\phi|^2+|\psi|^2)\phi=0.
\end{align}
It possesses the two-parameter soliton solution \cite{note-NLS} 
\begin{equation}\label{soliton-manakov}
\psi=e^{iz}\sin\theta\,\sech(x-\eta),\ \phi=e^{iz}\cos\theta\,\sech(x-\eta).
\end{equation}
Subject then, on the semi-infinite strip $x>0$ and $z\in[0,L]$, to the
boundary condition
\begin{equation}\label{bound-manakov}
  \psi(0,z)=Ae^{iz},\quad \phi(0,z)=Be^{iz},
\end{equation}
and vanishing values as $x\to\infty$, the Manakov system possesses 
solution (\ref{soliton-manakov}) provided the parameters $(\eta,\,\theta)$ 
are related to the driving amplitudes $A$ and $B$ by
\begin{equation}
 A=\sin\theta\,\sech\eta,\quad B=\cos\theta\,\sech\eta\ .
\end{equation}
The threshold curve in the $(A,B)$ plane is then obtained by solving
$J(\eta,\theta)=0$ as given by (\ref{BGS_NST_condition}).
The solution is $\eta=0$, that is, in the variables $(A,B)$, the circle
$A^2+B^2=1$. This is illustrated on Fig.\ref{fig:2} where we display a typical
soliton formation and the NST threshold curve for which the points represent
results of numerical simulations \cite{note-simuls}.

Then one can check that the threshold manifold of the $M$-component
Manakov system, $i\partial_z \phi_m+\partial^2_{x}\phi_m+2(\sum_{1}^M
|\phi_n|^2)\phi_m=0$, subject to the boundary condition $\phi_m(0,z)=A_m
e^{iz}$, is the $M$-dimensional sphere $\sum_{1}^M A_m^2=1$.

\paragraph*{Parametric instability and supratransmission.}

As can be seen from Figs.\ref{fig:1} and \ref{fig:2}, as soon as one of the
amplitudes $A$ or $B$ crosses the threshold curve, the evanescent wave profile
(soliton tail in integrable models) ceases to exist and gap solitons are emitted
by the driven boundary. It turns out that, at least in the models we have
investigated, the NST threshold manifold also corresponds to 
values of the parameters around which the stability of the solutions
changes. To substantiate this claim, we investigate why the conditions for
parametric and NST instabilities might indeed be the same by performing linear
stability analysis of the stationary solutions (\ref{psi-phi}) that we write
in the form
\begin{align} 
 & \psi(x,z)=(u(x)+U(x,z))e^{2iz}, \label{eq-pert-psi} \\
 & \phi(x,z)=\sqrt{\alpha}\, (v(x)+V(x,z))e^{iz}. \label{eq-pert-phi}
\end{align}
where $U$ and $V$ are small perturbations that satisfy, according to
(\ref{D3W-bound}),  the boundary condition
\begin{equation}\label{BCforUandV}
U(0,z)=0,\quad V(0,z)=0, 
\end{equation}
and vanishe at $x \rightarrow \infty$ for all $z$.
Defining the real-valued perturbation vectors $\Gamma_r=(U_r,V_r)^T$ and
$\Gamma_i=(U_i,V_i)^T$, where index $r$ ($i$) stands for real (imaginary) part,
linearization of system (\ref{D3W}) yields
\begin{equation} \label{eq-Gamma}
  \partial_z \Gamma_r + {\cal P}_- \Gamma_i = 0, \quad
  \partial_z \Gamma_i - {\cal P}_+ \Gamma_r = 0. 
\end{equation}
The matrix differential operators ${\cal P}_{\pm}$ are given by
\begin{equation}
 {\cal P}_{\pm} = 
 \left(\begin{array}{cc}\alpha & 0\\0 & 1\end{array}\right)
 \left(\begin{array}{cc}\frac{1}{2}\partial_x^2-2 & 2 v \\
  v  & \partial_x^2-1 \pm u   \end{array}\right).
\end{equation}
This is conveniently written as an eigenvalue problem by differentiating with
respect to $z$. For the real part $\Gamma_r(x,z)$ we may seek a
solution $\Gamma_r(x,z)=\Phi_\omega(x)\cos(\omega z)$ and obtain
\begin{equation} \label{EqGammar}
{\cal P}_- {\cal P}_+ \Phi_\omega = \omega^2  \Phi_\omega
\end{equation} 
with boundary conditions (use $\partial_z\Gamma_i(0,z)=0$)
\begin{equation} \label{BCforGammar}
\Phi_\omega(0)=0,\quad ({\cal P}_+ \Phi_\omega)(0) = 0.
\end{equation}
As ${\cal P}_{\pm}$ and $\Phi_\omega$ are real valued, $\omega^2$ is also real
valued. Then the solution is linearly stable when $\omega^2>0$ and unstable for
$\omega^2<0$, so that marginal instability is reached at the bifurcation point
$\omega=0$ providing the  parametric instability threshold.

An essential property of the operator ${\cal P}_+$, obtained by differentiation
of (\ref{eq-uv}) with respect to the parameters is
\begin{equation}
 {\cal P}_+ \frac{\partial}{\partial\eta}
 \left(\begin{array}{c}u\\ v\end{array}\right)=0,\quad
{\cal P}_+ \frac{\partial}{\partial\sigma}
 \left(\begin{array}{c}u\\ v\end{array}\right)=0.
\end{equation}
Thus a two-parameter family of solutions of
(\ref{EqGammar}) at $\omega=0$  reads
\begin{equation} \label{SolGammar}
\Phi_0(x)= a \frac{\partial}{\partial\eta} 
\left(\begin{array}{c}u\\ v\end{array}\right)
+ b \frac{\partial}{\partial\sigma}
\left(\begin{array}{c}u\\ v\end{array}\right),
\end{equation} 
for arbitrary constants $(a,b) \in \R^2$. Though it is not the most general
solution of (\ref{EqGammar}), it seems to be the only one able to satisfy the
boundary conditions (\ref{BCforGammar}). Requiring then $\Phi_0(0)=0$, with
$u(0)=A$ and $v(0)=B$, eventually yields $J(\eta,\sigma)=0$, namely the
parametric instability condition (\ref{BGS_NST_condition}). Thus the NST
threshold condition (\ref{BGS_NST_condition}), i.e., the condition for the
maximum allowed amplitudes $A$ and $B$ at the boundary $x=0$, actually coincides
with the onset of instability of the solution for the corresponding critical
values of the parameters (here $\eta$ and $\sigma$).

\begin{figure}[ht] 
\centerline{\epsfig{file=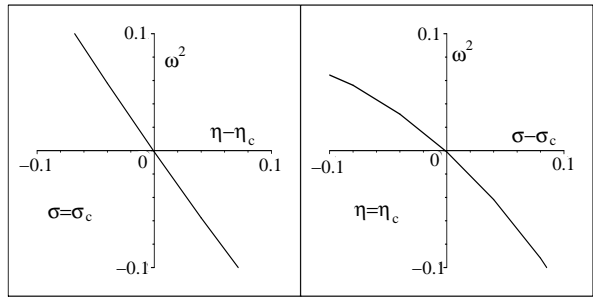,width=0.9\linewidth}}
\caption{Plots of the eigenvalue $\omega^2$ in terms of the
parameters $(\eta,\,\sigma)$ as a function of $\eta-\eta_c$ for
$\sigma=\sigma_c$ (left) and function of $\sigma-\sigma_c$ for
$\eta=\eta_c$ (right) again with $\alpha=1.1$.}\label{fig:3}
\end{figure}

To check this statement, we have computed numerically the eigenvalue $\omega^2$
of the differential equation (\ref{EqGammar}) around the bifurcation point
$\omega=0$ by varying  $\eta$ and $\sigma$ around their critical values $\eta_c$
and $\sigma_c$ defined by the solution of $J(\eta,\sigma)=0$. The result is
plotted in Fig.\ref{fig:3} where we have used the asymptotic series solution
$(u,v)$ at order $N=16$ as previously. As can be seen from the figure marginal
instability is actually reached at the criticality ($\eta_c$ and $\sigma_c$)
when $\omega$ crosses zero.

\paragraph*{Comments and conclusion.}

The method presented here can be readily applied to the simple
case of the scalar nonlinear Schr\"odinger equation $i\partial_z
\psi+\partial^2_x\psi+2|\psi|^2\psi=0$ to get interesting insight about the
occurrence of NST. It is found that (i) the asymptotic series solution actually
sums up exactly to furnish the one-soliton solution, (ii) the fundamental
parameter is the position $\eta$ of the soliton maximum, (iii) the threshold is
indeed the maximum amplitude of this static soliton reached at $\eta=0$, (iv)
the variations of the eigenvalue $\omega^2$ around zero is given by
$\omega^2=-4\eta+o(\eta)$ that, straightforwardly, gives the marginal
instability threshold $\eta=0$.

In such single component systems as NLS, the instability occurs always at the
maximum amplitude of the solution \cite{note-max}. On the contrary,  the
solution of the D3W model does not display any maximum nor any other
geometric evidence that the NST threshold has been reached.

In conclusion, we have solved the challenging problem of determining the
threshold for nonlinear supratransmission in nonintegrable $N$-component
systems. This is obtained in two steps: first by deriving an asymptotic solution
based on their linear evanescent profile that depends on $N$ parameters, and
second by solving Eq. (\ref{General_NST_condition}). In the parameter space
the latter condition results in a $N-1$ dimensional manifold that  determines
the change of stability of the (asymptotic) solution. Expressed in terms of the
amplitudes, it gives rise to the sought NST threshold. The situation is highly
simplified in the case of an integrable system, or a system that possesses an
exact static solitonlike solution, since one can work directly with the
solution to obtain the threshold.

Finally, since no analytical expression is required, the method can be
successfully applied to a wide class of nonintegrable nonlinear multicomponent
systems.

\begin{acknowledgments}
Work done as part of the programme GDR 3073 PhoNoMi2 (\textit{Photonique
Nonlin\'eaire et Milieux Micro\-structur\'es}).
\end{acknowledgments}


\begin{thebibliography}{a9}



\bibitem{nst-prl} 
F. Geniet, J. Leon, Phys Rev Lett 89 (2002) 134102
\bibitem{aubry}
P. Maniadis, G. Kopidakis, S. Aubry, Physica D 216 (2006) 121
\bibitem{daz}
J. E. Macias-Daz, A. Puri, Phys Lett A 366 (2007) 447
\bibitem{nst-bragg}
J. Leon, A. Spire, Phys Lett A 327 (2004) 474
\bibitem{taverner}
D. Taverner, N.G.R. Broderick, D.J. Richardson, R.I. Laming, M. Ibsen, 
Opt Lett 23 (1998) 328
\bibitem{ramaz1} 
R. Khomeriki, Phys Rev Lett 92 (2004) 063905
\bibitem{jl}
J. Leon, Phys Rev E 70 (2004) 056604
\bibitem{instab} 
J. Leon, Phys Lett A 319 (2003) 130
\bibitem{susanto} 
H. Susanto, SIAM J Appl Math 69 (2008) 111
\bibitem{ramaz2} 
R. Khomeriki, J. Leon, Phys Rev Lett 94 (2005) 243902
\bibitem{buryak} 
See e.g. the review:  A.V. Buryak, P. Di Trapani, D.V.
Skryabin, S. Trillo, Phys. Rep. 370 (2002) 63-235
\bibitem{majid}
V.A. Brazhnyi, V.V. Konotop, S. Coulibaly, M. Taki,
CHAOS 17 (2007) 037111
\bibitem{note-simuls} 
The points are obtained as follows: at fixed $B$  no
soliton is emitted for some $A$ while at $A+\epsilon$ solitons are generated.
The point is then set as $(A+\epsilon/2,B)$ and $\epsilon$ can be made as small
as needed within the accuracy of the numerical code. As component profiles are
initially set to zero, we smoothly increase amplitudes A and B from zero up to
their respective assigned value to avoid shocks. This is why simulations are
only displayed from $z = 60$ on.
\bibitem{note-BGS} 
The driving wavenumber is scaled to any value by using the invariance
D3W under the transformation $z\to k z$, $x\to\sqrt{k}\,x$ and
$(\psi,\phi,\delta)\to (\psi,\phi,\delta)/k$.
\bibitem{leo-assanto} 
G. Leo, G. Assanto, Opt. Lett. 22 (1997) 1391
\bibitem{canva}
M.T.G. Canva, R.A. Fuerst, S. Baboiu, G.I. Stegeman, G. Assanto,
Opt. Lett. 22 (1997) 1683
\bibitem{manakov}
S.V. Manakov, Sov Phys JETP 38 (1974) 248
\bibitem{note-NLS} 
The driving wavenumber is scaled to $k=1$ by using the invariance of
the system under the transformation $z\to k z$, $x\to\sqrt{k}\,x$ and
$(\psi,\phi)\to(\psi,\phi)/\sqrt{k}$.
\bibitem{note-max} 
The NST condition (\ref{General_NST_condition}) for a one-component system
leads to $\partial A/\partial\eta=0$ that is to 
$(\partial u/\partial x)(x=0)=0$.

 

\end{thebibliography}
\end{document}